\documentclass[aps,pra,amssymb,superscriptaddress,10pt,showpacs]{revtex4}
\usepackage{graphicx}
\usepackage{dcolumn}
\usepackage{bm}
\usepackage{color}
\usepackage{subfigure}

\begin{document}

\newcommand{\beq}{\begin{equation}}
\newcommand{\eeq}{\end{equation}}
\newcommand{\barr}{\begin{eqnarray}}
\newcommand{\earr}{\end{eqnarray}}

\def\bra#1{\langle{#1}|}
\def\ket#1{|{#1}\rangle}
\def\sinc{\mathop{\text{sinc}}\nolimits}
\def\cV{\mathcal{V}}
\def\cH{\mathcal{H}}
\def\cT{\mathcal{T}}
\renewcommand{\Re}{\mathop{\text{Re}}\nolimits}
\newcommand{\tr}{\mathop{\text{Tr}}\nolimits}

\definecolor{dgreen}{rgb}{0,0.5,0}
\newcommand{\green}{\color{dgreen}}
\newcommand{\RED}[1]{{\color{red}#1}}
\newcommand{\BLUE}[1]{{\color{blue}#1}}
\newcommand{\GREEN}[1]{{\color{dgreen}#1}}
\newcommand{\REV}[1]{{\color{red}[[#1]]}}
\newcommand{\KY}[1]{\textbf{\color{red}[[#1]]}}
\newcommand{\SP}[1]{{\color{blue} [[#1. Saverio]]}}
\newcommand{\rev}[1]{{\color{red}[[#1]]}}

\def\HN#1{{\color{magenta}#1}}
\def\DEL#1{{\color{yellow}#1}}

\title{Interference in a two-mode Bose system with $N$ particles is typical}

\author{Paolo Facchi}
\affiliation{Dipartimento di Fisica and MECENAS, Universit\`a di Bari, I-70126 Bari, Italy}
\affiliation{INFN, Sezione di Bari, I-70126 Bari, Italy}

\author{Hiromichi Nakazato}
\affiliation{Department of Physics, Waseda University, Tokyo 169-8555, Japan}

\author{Saverio Pascazio}
\affiliation{Dipartimento di Fisica and MECENAS, Universit\`a di Bari, I-70126 Bari, Italy}
\affiliation{INFN, Sezione di Bari, I-70126 Bari, Italy}

\author{Francesco V. Pepe}
\affiliation{Dipartimento di Fisica and MECENAS, Universit\`a di Bari, I-70126 Bari, Italy}
\affiliation{INFN, Sezione di Bari, I-70126 Bari, Italy}

\author{Kazuya Yuasa}
\affiliation{Department of Physics, Waseda University, Tokyo 169-8555, Japan}

\begin{abstract}
When a Bose-Einstein condensate is divided into
two parts, that are subsequently released and overlap,
interference fringes are observed. We show here that this
interference is typical, in the sense that most wave functions of
the condensate, randomly sampled out of a suitable ensemble, display interference. We make no
hypothesis of decoherence between the two parts of the
condensates.
\end{abstract}


\maketitle

\section{Introduction}
\label{sec-intro}

The experimental observation of Bose-Einstein condensation raised a number of deep questions regarding the phase of the condensate and its operational meaning,
its interference properties and other quantum concepts \cite{PW,SB,Leggett,BDZ,PS,PeSm,Leggettbook}. Interference fringes are observed when two
\textit{independently} prepared Bose-Einstein condensates (BECs) are
released and overlap \cite{exptBEC}. At first sight, this phenomenon seems to be
in contrast with commonly accepted wisdom on the Young-type
double-slit interference (first-order interference) experiments from independent sources: since the relative phase between the two wave functions is not
constant (this is what is meant by ``independent" sources), no
interference can be observed.
Notice however that the second-order
(two-particle) interference can be observed in a Hanbury-Brown and
Twiss type experiment \cite{HBT1,HBT}, where the quantum interference
 between the amplitudes corresponding to the two paths from the
independent sources to the detectors is observed \cite{feynman,ref:Loudon,ref:MandelWolf}.

Questions on the interference and the phase of the condensate have
attracted the attention of many researchers. This motivated a very
interesting debate on the most fundamental aspects of quantum
mechanics and the features of many-body quantum systems.
Javanainen and Yoo \cite{JY} proposed that the phase of the
condensate is established by measurement: an interference pattern is certainly observed in each \textit{single} experimental run, but the patterns shift from run to run, so that no interference persists if all the observed interference patterns are superimposed, with no contradiction with the common wisdom on the first-order interference.
The conclusion that the phenomenon can be ascribed to ``measurement-induced
interference" was also corroborated by related studies by Cirac
\emph{et~al.}\ \cite{CGNZ}, by Wong \emph{et~al.}\ \cite{WCW}, and by
Castin and Dalibard \cite{CD}. The main ideas are also explained
in review papers \cite{PW,Leggett,BDZ} and even in textbooks
\cite{PeSm,Leggettbook}, to signify the fundamental importance of
the problem and its correct interpretation. These results are also
interesting for the outlook they yield on symmetry breaking
phenomena \cite{SB,PS,Leggettbook,ref:LeggettSols}, and it is proposed that the fluctuations in the interference patterns can be exploited to probe interesting characteristics of many-body systems \cite{ref:NoiseCorr-AltmanDemlerLukin,ref:PolkovnikovAltmanDemlerPNAS,ref:GritsevAltmanDemlerPolkovnikov-NaturePhys2,ref:PolkovnikovEPL,ref:ImambekovGritsevDemlerVarenna,ref:GritsevDemlerPolkovnikov-PRA,ref:Hadzibabic-BECIntArray,ref:Hadzibabic-BECIntPhaseDefects,ref:Hadzibabic-BKT-Nature,ref:Hadzibabic2DTc,Raz}.

In this article, we would like to contribute to these issues. We
will show that two-particle interference is \emph{typical}, in the
sense that most wave functions, randomly sampled out of a suitable
subspace in the Hilbert space that describes the Bose-Einstein
condensate, display interference. No matter how the two gases are
prepared, or no matter how the splitting process of the atomic gas
into two parts takes place, one will typically observe an
interference pattern in each experimental run.

\section{Rationale}
\label{sec-razio}
In quantum mechanics one computes expectation values of
observables. The textbook interpretation is statistical: an
observable is represented by a self-adjoint operator and its expectation value
over a state yields the average that one obtains by repeating the
same experiment many times (with the system prepared in the same
state).

In the case to be examined here, one faces a difficulty: we are
required to discuss a ``single-shot'' experiment. When two
independently prepared condensates are released and overlap, the
two clouds will always interfere, in each experimental run, but
the offset of the interference pattern will change randomly from
run to run. On average (over many experimental runs) interference
is smeared out. The salient feature of each individual run is the
very presence of a clear interference pattern, with its
characteristics  (e.g., the distance between adjacent maxima). Can
one extract this single-run behavior from the quantum mechanical
formalism?

In this article we will argue that interference is robust with
respect to the state preparation. The main idea is the following. Each
time two condensates are experimentally prepared
(out of a single condensate, e.g.\ by inserting a ``wall'' between
them \cite{exptBEC,Schmied,Schmied2,Schmied3,Schmied4,Schmied5}), their wave function is sampled out of a given set. This
set is a portion of the total Hilbert space and presumably depends
on experimental procedures and details (state preparation). We
have no access to this information. We will therefore analyze the
\emph{typical} features of such a wave function, namely those
features that characterize its behavior and properties in the
overwhelming majority of cases. We will see that interference is
one of these distinctive features. Even though the offset
(positions of the maxima and minima) of the interference pattern
is random, so that interference will vanish on average over many
repetitions of the experimental run, the very \emph{presence} of an
interference pattern will emerge as a typical feature of the wave
function.

Clearly, no prediction is possible, in the quantum mechanical
formalism, if an expectation value is not computed. One must
always sandwich an operator between a bra and a ket at the end of
one's calculation. The only exception to this very general rule is
for a nonfluctuating observable, that is when the system is in an
eigenstate of the observable. In such a case the results of
single-shot experiments do not change from run to run. We will
show that asymptotically, i.e.\ for large number of particles, the
constitutive features of the interference pattern (period, intensity of the peaks, and so on)  do
not depend on the fact that a quantum expectation value is
computed, nor on the details of the preparation of the state.
Before we embark in the technical details of the calculation, we
anticipate that no hypothesis of decoherence between the two
independent condensates will prove necessary. The two condensates
will always be described by a randomly sampled pure state.

\section{Distribution of initial states}
\label{sec-typicality} In a typical experiment of BEC
interferometry, a condensate made up of $N$ bosons is distributed
among two orthogonal modes, $\psi_a(\bm{r})$ and $\psi_b(\bm{r})$.
If the modes are spatially separated at the initial time, they are
eventually let expand, overlap, and (possibly) interfere.

We assume that the
total number of bosons $N$ is fixed in the experiment. A useful
basis for a system of $N$ bosons is given by two-mode Fock states
\begin{equation}
 \label{fock} \ket{ \ell } := \left|
\left( \frac{N}{2}+\ell \right)_a, \left( \frac{N}{2}-\ell
\right)_b \right\rangle,
\end{equation}
in which the two modes, labelled with $a$ and $b$, have
well-defined occupation numbers. (We assume that $N$ is even for
simplicity.) In the second-quantization formalism, Fock states are
obtained by applying a sequence of creation operators to the
vacuum state $\ket{\Omega}$:
\begin{equation}
\ket{\ell} = \frac{1}{\sqrt{(N/2+\ell)! (N/2-\ell)!}}
({\hat{a}}^{\dagger})^{N/2+\ell}
({\hat{b}}^{\dagger})^{N/2-\ell} \ket{\Omega} .
\end{equation}
Due to the orthonormality of $\psi_a(\bm{r})$ and $\psi_b(\bm{r})$, the mode operators
\begin{equation}
\hat{a} = \int d\bm{r}\, \psi_a^*(\bm{r}) \hat{\Psi}(\bm{r}), \qquad
\hat{b} = \int d\bm{r}\, \psi_b^*(\bm{r}) \hat{\Psi}(\bm{r})
\end{equation}
satisfy the canonical commutation relations
\begin{equation}
[ \hat{a}, \hat{a}^{\dagger} ] =  [ \hat{b},
\hat{b}^{\dagger} ] = 1,
\end{equation}
and all the operators of mode $a$ commute with those of mode $b$.
Here $\hat{\Psi}(\bm{r})$ is the bosonic field operator, with the canonical commutation relations
\begin{equation}
[\hat{\Psi}(\bm{r}), \hat{\Psi}^\dagger(\bm{r'})]=\delta(\bm{r}-\bm{r}').
\end{equation}
The number
operators $\hat{N}_a={\hat{a}}^{\dagger} \hat{a}$ and
$\hat{N}_b={\hat{b}}^{\dagger} \hat{b}$ count the numbers of
particles in the two modes. We assume that, in each experimental
run, the initial state of the two-mode system is randomly picked
from the  subspace spanned by the Fock states
\begin{equation}
\mathcal{H}_n=\mathrm{span} \{ \ket{\ell} | {-n/2} < \ell < n/2\},
\label{6}
\end{equation}
with $0<n\leq N+1$, assuming that $n$ is odd.
The case $n=1$ represents the setup studied previously \cite{JY,CGNZ,WCW,CD,ref:PolkovnikovEPL,Paraoanu,YI}, while we are  interested here in the large $n$ case.
We will have in mind
the case $n = \mathrm{O}(\sqrt{N})$, but we will work in full
generality, with arbitrary $n$, and show that the following results are robust against larger scalings of $n$. (The case $n=N+1$  coincides
with uniform sampling over the full Hilbert space.) The assumption
of \emph{uniform} sampling is a simplifying one: the number of
states that are actually involved in the description and their
amplitude will depend on the experimental procedure and the way
the two BEC clouds are created \cite{esteve}.

If the pure state $\ket{\Phi_N}$ of the system is expanded in the Fock basis (\ref{fock}),
\begin{equation}
 \ket{\Phi_N} = \sum_{\ell=-N/2}^{N/2} z_{\ell}
\ket{\ell},
\end{equation}
the only states with nonvanishing probability will be the ones
with $-n/2 < \ell < n/2$, i.e.,
\begin{equation}
z_{\ell} = 0 \qquad \text{for} \qquad |\ell|> n/2.
\end{equation}
The coefficients $\{z_\ell\}$ are randomly sampled from the uniform distribution on the surface of the $2n$-dimensional unit sphere $\sum_\ell |z_{\ell}|^2=1$.
Due to this assumption of uniform sampling, the average
square modulus of the coefficients, $|z_{\ell}|^2$, is the
inverse of the subspace dimension, while the average of the
coefficients themselves, as well as all the quantities that depend
on relative phases in the superposition vanish:
\begin{equation}
\overline{z_{\ell_1}^* z_{\ell_2}} = \frac{1}{n}
\delta_{\ell_1,\ell_2},
\end{equation}
where we shall denote with a bar the statistical
average over the distribution of the coefficients of the state.
Notice that this quantity yields the density matrix associated with the uniform ensemble of $\ket{\Phi_N}$:
\begin{equation}
\hat{\rho}_N = \overline{\ket{\Phi_N}\bra{\Phi_N}} = \sum_{\ell_1,
\ell_2} \overline{z_{\ell_1} z^*_{\ell_2}} \ket{\ell_1}
\bra{\ell_2} = \frac{1}{n} \sum_{-n/2 < \ell < n/2 } \ket{\ell}
\bra{\ell} =: \frac{1}{n} \hat{P}_n ,
\end{equation}
where $\hat{P}_n$ is the projection onto the subspace $\mathcal{H}_n$.

The statistical average of the expectation value $A= \bra{\Phi_N}
\hat{A} \ket{\Phi_N}$ of observable $\hat{A}$ over the
ensemble of  states $\ket{\Phi_N}$ is
\begin{equation}\label{expGEN}
\overline{A} := \overline{ \bra{\Phi_N} \hat{A} \ket{\Phi_N}} =
\tr (\overline{ \ket{\Phi_N} \bra{\Phi_N}}\hat{A} ) =
\tr (\hat{\rho}_N \hat{A} ) = \frac{1}{n} \sum_{-n/2 < \ell < n/2
} \bra{\ell} \hat{A} \ket{\ell}.
\end{equation}
Its statistical variance, which measures the fluctuations of the
expectation value $A$ among the states $\ket{\Phi_N}$ of the
ensemble, reads
\begin{equation}\label{1}
(\delta A)^2 := \overline{ A^2 } - \overline{A}^2 = \overline{
\bra{\Phi_N} \hat{A} \ket{\Phi_N}^2} - \overline{ \bra{\Phi_N}
\hat{A} \ket{\Phi_N}}^2,
\end{equation}
and involves a quartic average [see \cite{cumulants}, Eq.\ (53)]
\begin{equation}\label{qave}
\overline{z_{\ell_1}^* z_{\ell_2}^* z_{\ell_3}
z_{\ell_4}} = \frac{1}{n(n+1)} \left(
\delta_{\ell_1,\ell_3} \delta_{\ell_2,\ell_4} +
\delta_{\ell_1,\ell_4} \delta_{\ell_2,\ell_3} \right).
\end{equation}
The quartic average is relevant also in the statistical average of
the variance of observable $\hat{A}$ in state $\ket{\Phi_N}$,
\begin{equation}\label{2}
\overline{( \Delta A)^2 }  :=
\overline{ \bra{\Phi_N} \hat{A}^2 \ket{\Phi_N}} -
\overline{ \bra{\Phi_N} \hat{A} \ket{\Phi_N}^2},
\end{equation}
which measures the average quantum fluctuations. Indeed, $(\Delta
A)^2$ is the expectation value of the observable $(\Delta
\hat{A})^2 = (\hat{A}-A)^2$.

Notice that by adding the two uncertainties $(\delta A)^2$  and
$\overline{ (\Delta A)^2 }$ one gets
\begin{equation}\label{fluctGEN}
 \overline{ (\Delta A)^2 } + (\delta A)^2 = \overline{ \bra{\Phi_N} \hat{A}^2 \ket{\Phi_N}} - \overline{ \bra{\Phi_N} \hat{A} \ket{\Phi_N}}^2 = \tr (\hat{\rho}_N \hat{A}^2 ) - \{\tr (\hat{\rho}_N \hat{A} )\}^2,
\end{equation}
which is nothing but the quantum variance of observable $\hat{A}$
at the mixed state $\hat{\rho}_N$. If one samples the initial
state from a degenerate distribution with $n=1$, in which only the
Fock state with equal number of particles in the two modes has
nonvanishing probability, the contribution to such a variance will
come only from the quantum fluctuations  $\overline{ (\Delta A)^2
}=  (\Delta A)^2$ at $\ket{\ell=0}$, while,
obviously, $(\delta A)^2=0$. On the other hand, if the ensemble is
made up of eigenstates of the observable $\hat{A}$, then the
quantum fluctuations vanish, $\overline{ (\Delta A)^2 }=0$, and
the only contribution comes from the statistical fluctuations
$(\delta A)^2$.

In general, there will be an interplay in (\ref{fluctGEN}), which
will depend on $n$, between the two sources of fluctuations. An
observed property is typical if $(\delta A)^2 \approx 0$, and is
run-independent if  $\overline{ (\Delta A)^2 } \approx 0$. Our
goal is to show that the period of the interference pattern can be
viewed and analyzed in a similar fashion.

\section{Average density and its Fourier transform}
Since we are interested in the quantities that are related to
interference, we will focus on those observables associated with
the spatial distribution of particles and their Fourier transforms \cite{ref:PolkovnikovAltmanDemlerPNAS,ref:GritsevAltmanDemlerPolkovnikov-NaturePhys2,ref:PolkovnikovEPL,ref:ImambekovGritsevDemlerVarenna,YI,AYI}.
In this section we will introduce the relevant averages in the
general case, postponing quantitative considerations on
interference to the following section. In the second-quantization
formalism, the spatial density observable is given by the operator
\begin{equation}
\hat{\rho}(\bm{r}) = \hat{\Psi}^{\dagger}(\bm{r})
\hat{\Psi}(\bm{r}),
\end{equation}
while its Fourier transform reads
\begin{equation}
\widehat{\tilde{\rho}}(\bm{k}) := \mathcal{F}
[\hat{\rho}](\bm{k}) = \int d\bm{r} \, e^{-i \bm{k}
\cdot \bm{r} } \hat{\rho}(\bm{r}).
\end{equation}
Expanding the field operators and taking the  expectation value
(\ref{expGEN}), one finds that the average density
\begin{equation}\label{exprho}
 \overline{\rho(\bm{r})} = \overline{ \bra{\Phi_N}
\hat{\rho}(\bm{r}) \ket{\Phi_N}}  = \frac{N}{2} \,\Bigl(
\rho_a(\bm{r}) + \rho_b(\bm{r}) \Bigr), \quad \text{with} \quad
\rho_{a,b}(\bm{r}) := |\psi_{a,b}(\bm{r})|^2,
\end{equation}
is merely the sum of the particle densities in the two modes, with
no interference between them. Clearly, this property holds also
for the Fourier transform. This result apparently contrasts with
experiment, as interference is observed even if no phase coherence
between the particles in the two modes is present. However, as we
will show in the following, the average (\ref{exprho})
cannot give sufficient information on the result of a {\it single}
experimental run, since its fluctuations can be very large.

On the other hand, we will show that the outcome of a single
run can be inferred, within a controlled degree of
approximation, from the study of a different operator, namely \cite{ref:PolkovnikovEPL,YI,AYI}
\begin{equation}\label{rho2k}
 \hat{R}(\bm{k}) :=
\widehat{\tilde{\rho}}^\dag(\bm{k})\widehat{\tilde{\rho}}(\bm{k})
= \widehat{\tilde{\rho}}(\bm{-k}) \widehat{\tilde{\rho}}(\bm{k}) =
\int d\bm{r}\, d\bm{r}' \,e^{-i\bm{k}\cdot(\bm{r}-\bm{r}')}
\hat{\Psi}^{\dagger}(\bm{r}) \hat{\Psi}^{\dagger}(\bm{r}')
\hat{\Psi}(\bm{r}') \hat{\Psi}(\bm{r}) + \int
d\bm{r}\,\hat\rho(\bm{r}) =: \hat{r}(\bm{k}) + \hat{N}  .
\end{equation}
 The normal ordering of the field operators in the
first integral incorporates the symmetry $\bm{k}\leftrightarrow -\bm{k}$. Notice that the expectation
value of the number of particles $\hat{N}$ is a constant for all
states $\ket{\Phi_N}$ and is thus immaterial in our study.
Under specific assumptions on the values of $N$ and $n$ in (\ref{6}), we will
show in this and the following sections that the fluctuations
around the average value
$\overline{R(\bm{k})}$ are negligible.

We start by expanding $\hat{r}(\bm{k})$ in mode
operators, which will be useful also in the computation of the
statistical and quantum fluctuations
\begin{eqnarray}\label{oprk}
\hat{r}(\bm{k}) & = & |\tilde{\rho}_a(\bm{k})|^2 \hat{N}_a
(\hat{N}_a-1) + |\tilde{\rho}_b(\bm{k})|^2 \hat{N}_b (\hat{N}_b-1)
\nonumber\\
&&{}+\Bigl(
\tilde{\rho}_a^*(\bm{k}) \tilde{\rho}_b(\bm{k}) +
\tilde{\rho}_b^*(\bm{k}) \tilde{\rho}_a(\bm{k}) +
|\mathcal{F}[ \psi_b^*\psi_a](\bm{k})|^2 +
|\mathcal{F}[\psi_a^*\psi_b](\bm{k})|^2
\Bigr)
\,\hat{N}_a\hat{N}_b
\nonumber \\
&&{} +
\Bigl[
\Bigl(
\tilde{\rho}_b(-\bm{k}) \mathcal{F}[
\psi_b^*\psi_a](\bm{k})
+
\tilde{\rho}_b(\bm{k})
\mathcal{F}[\psi_b^*\psi_a](-\bm{k}) \Bigr)\, \hat{N}_{b}{\hat{b}}^{\dagger}\hat{a}
 +  \Bigl(
 \mathcal{F}[
\psi_b^*\psi_a](-\bm{k})
 \tilde{\rho}_a(\bm{k})
 +
\mathcal{F} [ \psi_b^*\psi_a](\bm{k})
\tilde{\rho}_a(-\bm{k})
\Bigr)\,
{\hat{b}}^{\dagger}\hat{a}\hat{N}_{a}
\nonumber \\
&&\qquad{} +
 \mathcal{F}[ \psi_b^*\psi_a](-\bm{k})
 \mathcal{F}[ \psi_b^*\psi_a](\bm{k})
 (\hat{b}^{\dagger})^2
{\hat{a}}^2 + \text{h.c.}
 \Bigr]
 + \text{other modes}.
\end{eqnarray}
The average squared Fourier component of the density can be
obtained after taking the statistical averages of the mode
operators. Due to the uniform sampling on the relevant subspace, only
operators with diagonal matrix elements in the Fock basis yield
nonvanishing contributions:
\begin{eqnarray}\label{sums2}
S_{2,0} & = & \frac{1}{n} \sum_{-n/2 < \ell < n/2 } \bra{\ell}
\hat{N}_{a,b}(\hat{N}_{a,b}-1) \ket{\ell} = \frac{N^2-2N}{4} +
\frac{n^2-1}{12}, \\
S_{1,1} & = &\frac{1}{n} \sum_{-n/2 < \ell < n/2 } \bra{\ell}
\hat{N}_a \hat{N}_b \ket{\ell} = \frac{N^2}{4} - \frac{n^2-1}{12}.
\end{eqnarray}
The final result reads
\begin{equation}
\overline{R(\bm{k})} = N +
\Bigl(
|\tilde{\rho}_a(\bm{k})|^2 +
|\tilde{\rho}_b(\bm{k})|^2
\Bigr)\, S_{2,0}
+ \Bigl(
\tilde{\rho}_a^*(\bm{k}) \tilde{\rho}_b(\bm{k}) +
\tilde{\rho}_b^*(\bm{k}) \tilde{\rho}_a(\bm{k}) +
|\mathcal{F}[\psi_a^*\psi_b](-\bm{k})|^2 +
|\mathcal{F}[ \psi_b^*\psi_a](-\bm{k})|^2
\Bigr)\,
S_{1,1}. \label{rho2kfinal}
\end{equation}
 In order to estimate the fluctuations of
$R(\bm{k})=\bra{\Phi_N}\hat{R}(\bm{k})\ket{\Phi_N}$ around its
average and prove that they are small in the large-$n$ limit, we
shall consider the covariance
\begin{equation}\label{enscov}
(\delta R)^2 (\bm{k},\bm{k}') := \overline{ R(\bm{k}) R(\bm{k}') }
- \overline{R(\bm{k})} \cdot \overline{R(\bm{k}')} = \overline{
r(\bm{k}) r(\bm{k}') } - \overline{r(\bm{k})} \cdot
\overline{r(\bm{k}')}.
\end{equation}
The statistical average of the quartic product (\ref{qave}) and a
comparison with the general expression of $\overline{r(\bm{k})}$
yield
\begin{equation}
\overline{r(\bm{k}) r(\bm{k}')} = \frac{n}{n+1}
\overline{r(\bm{k})} \cdot \overline{r(\bm{k}')} +
\frac{1}{n(n+1)} \sum_{-n/2< \ell_1,\ell_2<n/2} \bra{\ell_1}
\hat{r}(\bm{k}) \ket{\ell_2} \bra{\ell_2} \hat{r}(\bm{k}')
\ket{\ell_1}.
\end{equation}
Thus, the covariance (\ref{enscov}) stems from two contributions
\begin{equation}
(\delta R)^2 (\bm{k},\bm{k}') = (\delta R_{\text{diag}})^2
(\bm{k},\bm{k}') + (\delta R_{\text{off}})^2 (\bm{k},\bm{k}'),
\end{equation}
where
\begin{equation}
(\delta R_{\text{diag}})^2 (\bm{k},\bm{k}') := - \frac{1}{n+1}
 \overline{r(\bm{k})} \cdot \overline{r(\bm{k}')} +
\frac{1}{n(n+1)} \sum_{\ell} \bra{\ell}
\hat{r}(\bm{k}) \ket{\ell} \bra{\ell} \hat{r}(\bm{k}') \ket{\ell}
\end{equation}
includes all the contributions coming from the diagonal matrix
elements in the two-mode Fock basis, while
\begin{equation}
(\delta R_{\text{off}})^2 (\bm{k},\bm{k}') := \frac{1}{n(n+1)}
\sum_{\ell_1 \neq \ell_2} \bra{\ell_1} \hat{r}(\bm{k})
\ket{\ell_2} \bra{\ell_2} \hat{r}(\bm{k}') \ket{\ell_1}
\end{equation}
contains the off-diagonal terms. Direct computation shows that the summation appearing in $\delta^2
R_{\text{diag}}$ cancels with the first term at the highest
order $(N^4/n)$, leaving
\begin{eqnarray}\label{covdiag}
(\delta R_{\text{diag}})^2 (\bm{k},\bm{k}') & = & \frac{N^2 n}{12}
\,\Bigl( |\tilde{\rho}_a(\bm{k})|^2 - |\tilde{\rho}_b(\bm{k})|^2
\Bigr) \,\Bigl( |\tilde{\rho}_a(\bm{k}')|^2 -
|\tilde{\rho}_b(\bm{k}')|^2 \Bigr) \nonumber \\ &&{} +
\frac{n^3}{180} \,\Bigl( \tilde{\rho}_a^*(\bm{k})
\tilde{\rho}_b(\bm{k}) + \tilde{\rho}_b^*(\bm{k})
\tilde{\rho}_a(\bm{k}) + |\mathcal{F}[ \psi_a^*\psi_b](-\bm{k})|^2
+ |\mathcal{F}[ \psi_b^*\psi_a](-\bm{k})|^2 \Bigr) \nonumber \\ &
& \qquad\quad {}\times\Bigl( \tilde{\rho}_a^*(\bm{k})
\tilde{\rho}_b(\bm{k}) + \tilde{\rho}_b^*(\bm{k})
\tilde{\rho}_a(\bm{k}) + |\mathcal{F}[ \psi_a^*\psi_b](-\bm{k})|^2
+ |\mathcal{F}[\psi_b^*\psi_a](-\bm{k})|^2 \Bigr) + \text{O}(N^2),
\end{eqnarray}
where we are assuming that terms of order $N^2$ can be neglected,
see the discussion in the following section. The highest-order
contribution to $(\delta R)^2$, coming from the off-diagonal part,
reads
\begin{eqnarray}
(\delta R_{\text{off}})^2 (\bm{k},\bm{k}') & = & \frac{N^4}{16 n}
\,\Bigl[ \mathcal{F}[\psi_b^*\psi_a](\bm{k})
\mathcal{F}[\psi_b^*\psi_a](-\bm{k})
\mathcal{F}[\psi_a^*\psi_b](\bm{k}')
\mathcal{F}[\psi_a^*\psi_b](-\bm{k}')
 \nonumber \\
&&\qquad\quad{} +\Bigl(
\mathcal{F}[\psi_b^*\psi_a](-\bm{k})
I_{ab}(\bm{k})
 +
 I_{ab}(-\bm{k})
\mathcal{F}[\psi_b^*\psi_a](\bm{k})
\Bigr) \nonumber \\
& &\qquad\qquad
{}\times \Bigl(
\mathcal{F}[\psi_a^*\psi_b](-\bm{k}')
I_{ab}(\bm{k}')
 + I_{ab}(-\bm{k}')
\mathcal{F}[\psi_a^*\psi_b](\bm{k}')
\Bigr) +
\text{c.c.}
\Bigr] + \text{O}\!\left( \frac{N^4}{n^2},n N^2
\right),
\label{covoff}
\end{eqnarray}
with
\begin{equation}
I_{ab}(\bm{k}) = \tilde{\rho}_a(\bm{k}) + \tilde{\rho}_b(\bm{k}).
\end{equation}
Thus, since $\overline{R(\bm{k})}=\text{O}(N^2)$, the relative
covariance of $R(\bm{k})$   vanishes like
$n^{-1}$ in the limit of large $n$.

In view of the following discussion on the typicality of
interference, let us also observe that, due to the form
(\ref{oprk}) of $\hat{r}(\bm{k})$, all the off-diagonal
contributions involve products of the type
$\mathcal{F}[\psi_a^*\psi_b](-\bm{k})\mathcal{F}[\psi_a^*\psi_b](\bm{k})$
or
$\mathcal{F}[\psi_a^*\psi_b](-\bm{k})\tilde{\rho}_{a,b}(\bm{k})$.
In the cases we will study below these terms will vanish due to the symmetric structure of the modes,
and thus $(\delta R_{\text{diag}})^2$ will be the only relevant
contribution to the covariance. Moreover, the structure of the higher order  contributions in~(\ref{covoff}) is the same and therefore the last term will also vanish.
This feature relaxes the condition for the vanishing relative
covariance of $R(\bm{k})$: it vanishes in the limit $N\to\infty$, irrespectively of $n$.

Another important quantity in the analysis of
interference patterns is the statistical average of the quantum
covariance of the observable $\hat{R}(\bm{k})$ around its
expectation value
\begin{eqnarray}\label{DeltaR2}
\overline{(\Delta R)^2 (\bm{k},\bm{k}')} &=& \overline{
\bra{\Phi_N} \hat{R}(\bm{k}) \hat{R}(\bm{k}') \ket{\Phi_N} } -
\overline{ \bra{\Phi_N} \hat{R}(\bm{k}) \ket{\Phi_N} \bra{\Phi_N}
\hat{R}(\bm{k}') \ket{\Phi_N} }
\nonumber\\
&=& \overline{ \bra{\Phi_N}
\hat{r}(\bm{k}) \hat{r}(\bm{k}') \ket{\Phi_N} } -
\overline{r(\bm{k})} \cdot \overline{r(\bm{k}')} - (\delta R)^2 (\bm{k},\bm{k}').
\end{eqnarray}
After normal-ordering the field operators,
the operator appearing in the first term of (\ref{DeltaR2}) can be
expressed as
\begin{eqnarray}\label{rho4k}
\hat{r}(\bm{k}) \hat{r}(\bm{k}') & = & \int d^3\bm{r}\, d^3\bm{r}' \,d^3 \bm{r}''\, d^3\bm{r}'''\, e^{-i \bm{k}\cdot(\bm{r}-\bm{r'})- i
\bm{k}'\cdot(\bm{r''}-\bm{r'''})} \hat{\Psi}^{\dagger}(\bm{r})
\hat{\Psi}^{\dagger}(\bm{r}') \hat{\Psi}^{\dagger}(\bm{r}'')
\hat{\Psi}^{\dagger}(\bm{r}''') \hat{\Psi}(\bm{r}''')
\hat{\Psi}(\bm{r}'') \hat{\Psi}(\bm{r}') \hat{\Psi}(\bm{r})
\nonumber \\ & &{}+ \int d^3\bm{r}\, d^3\bm{r}'\, d^3\bm{r}''\,
\hat{\Psi}^{\dagger}(\bm{r}) \hat{\Psi}^{\dagger}(\bm{r}')
\hat{\Psi}^{\dagger}(\bm{r}'') \hat{\Psi}(\bm{r}'')
\hat{\Psi}(\bm{r}') \hat{\Psi}(\bm{r}) \nonumber \\ & &
\qquad\qquad\qquad\qquad
{}\times \Bigl(
 e^{-i(\bm{k}+\bm{k}')\cdot\bm{r} + i
(\bm{k}\cdot\bm{r'} + \bm{k}'\cdot\bm{r''})} +
e^{-i(\bm{k}-\bm{k}')\cdot\bm{r} + i (\bm{k}\cdot\bm{r'} -
\bm{k}'\cdot\bm{r''})} + \text{c.c.}
\Bigr) + \text{O}(N^2).
\end{eqnarray}
The result for its averaged expectation value reads
\begin{equation}\label{deltarho}
\overline{ \bra{\Phi_N} \hat{r}(\bm{k}) \hat{r}(\bm{k}')
\ket{\Phi_N} } = F_{4,0}(\bm{k},\bm{k}') S_{4,0} +
F_{3,1}(\bm{k},\bm{k}') S_{3,1} + F_{2,2}(\bm{k},\bm{k}') S_{2,2}
+ F_{3,0}(\bm{k},\bm{k}') S_{3,0} + F_{2,1}(\bm{k},\bm{k}')
S_{2,1} + \text{O}(N^2),
\end{equation}
where the sums $S_{i,j}$ read
\begin{eqnarray}
S_{4,0} & = & \frac{1}{n} \sum_{-n/2 < \ell < n/2 } \bra{\ell}
\hat{N}_{a,b}(\hat{N}_{a,b}-1)(\hat{N}_{a,b}-2)(\hat{N}_{a,b}-3)
\ket{\ell} \nonumber \\
& = & \frac{N^4-12N^3}{16} + \frac{n^4}{80} + \frac{(N^2-6N)
n^2}{8} +
\text{O}(N^2), \\
S_{3,1} & = & \frac{1}{n} \sum_{-n/2 < \ell < n/2 } \bra{\ell}
\hat{N}_{b,a}\hat{N}_{a,b}(\hat{N}_{a,b}-1)(\hat{N}_{a,b}-2)
\ket{\ell}
\nonumber\\
&=& \frac{N^4- 6 N^3}{16} - \frac{n^4}{80} + \frac{N
n^2}{8} + \text{O}(N^2), \\
S_{2,2} & = & \frac{1}{n} \sum_{-n/2 < \ell < n/2 } \bra{\ell}
\hat{N}_{a}\hat{N}_{b}(\hat{N}_{a}-1)(\hat{N}_{b}-1) \ket{\ell}
\nonumber \\ & = & \frac{N^4- 4 N^3}{16} + \frac{n^4}{80} -
\frac{(N^2-2N) n^2}{24} + \text{O}(N^2), \\
S_{3,0} & = & \frac{1}{n} \sum_{-n/2 < \ell < n/2 } \bra{\ell}
\hat{N}_{a,b}(\hat{N}_{a,b}-1)(\hat{N}_{a,b}-2) \ket{\ell} =
\frac{N^3 + N n^2}{8} + \text{O}(N^2), \\
S_{2,1} & = & \frac{1}{n} \sum_{-n/2 < \ell < n/2 } \bra{\ell}
\hat{N}_{b,a}\hat{N}_{a,b}(\hat{N}_{a,b}-1) \ket{\ell} = \frac{3
N^3 - N n^2}{24} + \text{O}(N^2),
\end{eqnarray}
while $F_{i,j}(\bm{k},\bm{k}')$ are symmetric functions of their
arguments, whose form is given in the Appendix. In the following
section, we will analyze the two cases in which the distribution of
$R(\bm{k})$ displays the sharp peaks that provide the information on the
interference pattern in each experimental run, with fluctuations being
negligible in proper ranges of $N$ and $n$.

\section{Typicality of interference}

\subsection{Counterpropagating plane-wave modes}

In this section, we will discuss a paradigmatic example of
interference in Bose systems \cite{JY,CGNZ}, to show how and under
which physical conditions interference becomes typical. Let us
assume that the system be confined in a unit volume, with periodic
boundary conditions, and let us choose as orthogonal modes
two counterpropagating plane waves
\begin{equation}\label{planewaves}
\psi_a(\bm{r}) = e^{i \bm{k}_0 \cdot \bm{r}}, \qquad
\psi_b(\bm{r}) = e^{-i \bm{k}_0 \cdot \bm{r}},
\end{equation}
with $\bm{k}_0 \neq 0$. The building blocks of the average
value of  $R(\bm{k})$ in (\ref{rho2kfinal}) and its
covariance (\ref{rho4k}) are the Fourier series of the densities
and the products $\psi_a^*\psi_b$. In this case, the form
(\ref{planewaves}) of the wave functions yields the
Fourier coefficients
\begin{equation}\label{planewavesFT}
\tilde{\rho}_{a}(\bm{k})=\tilde{\rho}_b(\bm{k})=
\delta_{\bm{k},0}, \quad \mathcal{F}[ \psi_a^*\psi_b
](\bm{k}) = \delta_{\bm{k},-2\bm{k}_0}, \quad
\mathcal{F}[ \psi_b^*\psi_a](\bm{k}) =
\delta_{\bm{k},+2\bm{k}_0},
\end{equation}
which make all terms of the form $\tilde{\rho}_{a,b}(\bm{k})
\mathcal{F}[\psi_a^*\psi_b](\pm\bm{k})$ and $ \mathcal{F}[\psi_a^*\psi_b](-\bm{k})\mathcal{F}[
\psi_a^*\psi_b](\bm{k})$
in the covariances identically vanish (see the Appendix). The
average value (\ref{rho2kfinal}) of the squared Fourier
component of the density reads
\begin{equation}\label{rho2kPW}
\overline{R(\bm{k})} = N^2 \delta_{\bm{k},0} + \left(
\frac{N^2}{4} - \frac{n^2}{12} \right)(
\delta_{\bm{k},-2\bm{k}_0} + \delta_{\bm{k},+2\bm{k}_0}
) +
\text{O}(N).
\label{eqn:AveRplane}
\end{equation}
As we shall see, this quantity is essential to characterize the
interference pattern appearing in a single experimental run.
However, it is first necessary to compute its covariance in order
to estimate its fluctuations. Both $(\delta R_{\text{off}})^2$ and
the terms of $\text{O}(N^2 n)$ in $(\delta R_{\text{diag}})^2$
identically vanish. Thus, for large $n$, the dominant contribution
to $(\delta R)^2(\bm{k},\bm{k}')$ comes from the $\text{O}(n^3)$
terms in the diagonal part, otherwise the covariance is
$\text{O}(N^2)$:
\begin{equation}
(\delta R)^2 (\bm{k},\bm{k}') = \frac{n^3}{180} (
\delta_{\bm{k},-2\bm{k}_0} + \delta_{\bm{k},+2\bm{k}_0} ) (
 \delta_{\bm{k}',-2\bm{k}_0} + \delta_{\bm{k}',+2\bm{k}_0}
) + \text{O} (N^2).
\end{equation}

The use of (\ref{planewavesFT}) largely simplifies the functions
$F_{i,j}$ that appear in the statistical average of
the quantum covariance [see (\ref{deltarho})], that finally
reads
\begin{equation}\label{rho4kPW}
\overline{(\Delta R)^2 (\bm{k},\bm{k}')} = \left( \frac{N^3}{4} -
\frac{N n^2}{12} + \frac{n^4 - n^3}{180} \right) (
\delta_{\bm{k},-2\bm{k}_0} + \delta_{\bm{k},+2\bm{k}_0} ) (
 \delta_{\bm{k}',-2\bm{k}_0} + \delta_{\bm{k}',+2\bm{k}_0}
) + \text{O}(N^2).
\end{equation}
Important comments must be made on the results
(\ref{rho2kPW})--(\ref{rho4kPW}). First, the order of magnitude of
the covariance is smaller than the order $N^4$ of the squared
average value $\overline{R}^2$, unless $n=\text{O}(N)$. This
implies that if $n=\text{o} (N)$ (i.e., $n/N \to 0$ for $N\to\infty$) fluctuations around the average of
$\overline{R}$ in (\ref{rho2kPW}) are negligible, and the distribution
of its values is peaked around its most probable value.

Therefore, we can identify two different regimes: for $n=\text{O}(N^{3/4})$, the relative fluctuations $\Delta R/R$
scale as $N^{-1/2}$ and are of the same order as the quantum fluctuations of the central Fock state; for larger values of $n$, say $n\sim N^{\alpha}$ with $\alpha\geq 3/4$, but still
$\text{o}(N)$ (i.e.\ $\alpha <1$), the fluctuations scale as
$(n/N)^2\sim N^{-2(1-\alpha)}$. Notice also that in the latter
regime the order of magnitude of fluctuations is the same as the
subleading terms in (\ref{rho2kPW}), and that terms $N n^2$ and
$n^3$ are subdominant in all regimes.

If fluctuations on the peaks appearing in the average value
$\overline{R(\bm{k})}$ are small, we expect that in each run, for
any initial state sampled from our distribution, as long as $n=\mathrm{o}(N)$, the outcome of
the measurement of this quantity, which is obtained by
Fourier-transforming the experimental density, will read, with
overwhelming probability,
\begin{equation}
|\tilde{\rho}(\bm{k})|^2 = N^2
\left[ \delta_{\bm{k},0} + \frac{1}{4} (
\delta_{\bm{k},-2\bm{k}_0} + \delta_{\bm{k},+2\bm{k}_0}
) +
\text{O}\! \left( \frac{1}{\sqrt{N}}, \left( \frac{n}{N} \right)^2
\right) \right].
\end{equation}
This result, together with the normalization and condition of reality of the
density, determines the form of the Fourier transform of the density
\begin{equation}
\tilde{\rho}(\bm{k}) \approx N \left[
\delta_{\bm{k},0} + \frac{1}{2} ( \delta_{\bm{k},-2\bm{k}_0}
e^{-i \phi_{s} }+ \delta_{\bm{k},+2\bm{k}_0} e^{i
\phi_{s}} ) \right],
\end{equation}
which finally yields the single-run interference pattern
\begin{equation}
\rho(\bm{r}) \approx 2 N \cos^2
( \bm{k}_0 \cdot\bm{r}+\phi_{s}).
\end{equation}
Note that this perfect visibility is guaranteed by the condition $n=\mathrm{o}(N)$ in (\ref{eqn:AveRplane}).
The phase offset $\phi_s$ is fixed in each experimental run, but must fluctuate from run to run:
the density is indeed uniform on average
\begin{equation}
\overline{{\rho}(\bm{r})} = \overline{ 2 N \cos^2
( \bm{k}_0 \cdot\bm{r}+\phi_s)}= N.
\end{equation}
The randomness of the phase offset accounts for the fact that,
unlike $|\tilde{\rho}|^2$, fluctuations around the average
value of $\tilde{\rho}$ are very large.

\subsection{Expanding Gaussian modes}
\begin{figure}
\centering
\includegraphics[width=0.5\textwidth]{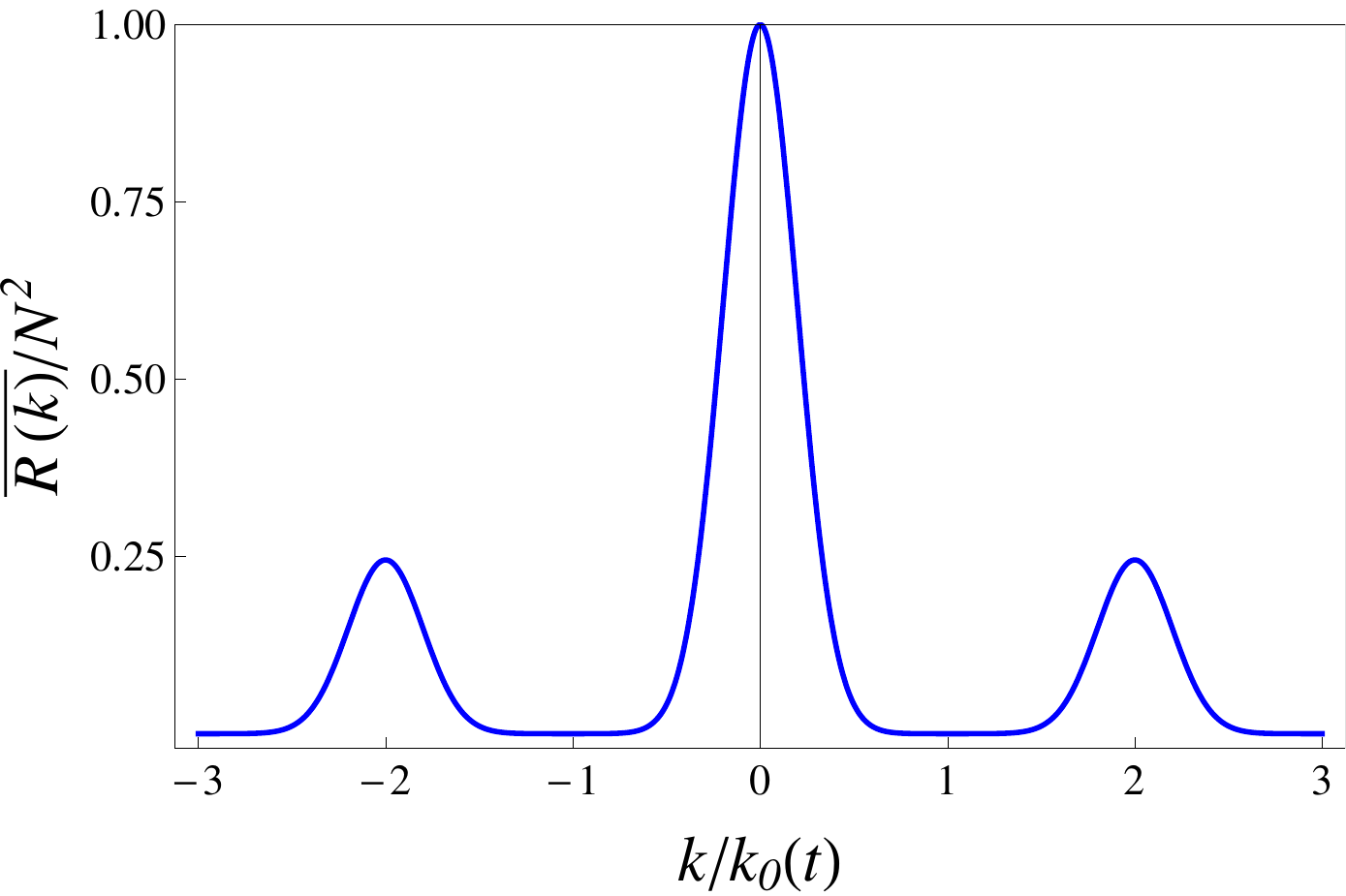}
\caption{The average value of $R(k)$ in the large $N$ limit, for two
expanding and overlapping Gaussian wave packets. The curve refers
to $\alpha=5$ and $t=10\alpha$. The width $1/t$ and the
distance $2\alpha/t$ among the three Gaussian functions decrease
with time. }\label{gaussianrho2}
\end{figure}

In this subsection we will analyze a more realistic model, describing
a physical situation that is closer to experimental
implementation. The cold atoms are initially trapped in two
Gaussian clouds by an external potential. The distance between the
centers of the distributions is larger than their widths, so that
the initial wave packets do not appreciably overlap. The trap is then released
and the clouds expand in free space until they overlap and
interfere. We will explicitly consider the time evolution of the
system in one spatial dimension, for simplicity: if the
scattering between the particles in the condensates is neglected,
the time dependence can be evaluated by observing that the
correlation functions at time $t$ can be obtained by replacing the
initial modes $\psi_{a,b}(x)$ with their time-evolved ones
\begin{equation}
\psi_{a,b}(x,t) = \exp\!\left( \frac{i \hbar t}{2 m }
\frac{\partial^2}{\partial x^2} \right) \psi_{a,b}(x).
\end{equation}
To simplify notation, lengths will be expressed in units of the
Gaussian mode width $\sigma$ and times in units
$m\sigma^2/\hbar$. The initial modes thus read
\begin{equation}
\psi_a(x) = \frac{1}{\pi^{1/4}} e^{- (x+\alpha)^2/2}, \qquad
\psi_b(x) = \frac{1}{\pi^{1/4}} e^{- (x-\alpha)^2/2},
\end{equation}
where $\alpha$ is half distance between the peaks of the
Gaussians and must be chosen large enough in order to ensure that
$\psi_{a,b}$ be quasi-orthogonal. Time evolution is most easily
expressed in the momentum basis (notice that $k$ is in units
$\sigma^{-1}$)
\begin{eqnarray}
\tilde{\psi}_a (k, t) & = & \int dx\, e^{-ikx} \psi_a (x, t) = \sqrt{2}\,
\pi^{1/4} e^{+ik\alpha} e^{-\frac{1}{2} (1+it) k^2}, \\
\tilde{\psi}_b (k, t) & = & \int dx\, e^{-ikx} \psi_b (x, t) = \sqrt{2}\,
\pi^{1/4} e^{-ik\alpha} e^{-\frac{1}{2} (1+it) k^2}.
\end{eqnarray}
The Fourier transforms relevant to the average value of $R(k)$ and
its covariance can be computed by convolution, yielding
\begin{eqnarray}
\tilde{\rho}_a (k,t) & = & \exp \!\left( -\frac{1}{4} (1+t^2) k^2+
i k \alpha \right) = \tilde{\rho}_b (-k,t), \\
\mathcal{F}[
\psi_a^* \psi_b ] (k,t) & = & \exp\!\left(-
\frac{\alpha^2}{1+t^2} \right) \exp\!\left(
 -\frac{1}{4} (1+t^2)
[ k + 2 k_0(t)]^2 \right) = \mathcal{F}[
\psi_b^* \psi_a](-k,t),
\end{eqnarray}
with
\begin{equation}
k_0(t) = \frac{\alpha t}{1+t^2}.
\end{equation}
These results can be used to evaluate the dominant contribution,
for large $N$ and $n=\text{o}(N)$, to the average value~(\ref{rho2kfinal}), in
the large-time limit ($t\gg 1,\alpha$)
\begin{eqnarray}\label{rho2kgauss}
\overline{ R(k,t) } & \approx & \frac{N^2}{4} \left[ \Bigl(
\tilde{\rho}_a(k,t)+\tilde{\rho}_b(k,t) \Bigr)^2 + \Bigl(
\mathcal{F}[\psi_a^* \psi_b] (k,t) \Bigr)^2 + \Bigl(
\mathcal{F}[\psi_b^* \psi_a](k,t) \Bigr)^2 \right]
\nonumber \\ & \approx & N^2 \left\{ \exp\! \left( - \frac{1}{2} t^2
k^2 \right)
+ \frac{1}{4} \left[ \exp \!\left( -
\frac{1}{2} t^2 [k+2k_0(t)]^2 \right) + \exp\! \left( - \frac{1}{2}
t^2 [k-2k_0(t)]^2 \right) \right] \right\}.
\end{eqnarray}
The average value, plotted in Fig.\ \ref{gaussianrho2}, is thus
characterized by three Gaussian functions in momentum space, whose
widths $1/t$ and distance $2\alpha/t$ decrease with time. The side
peaks at  $\pm 2 k_0(t)$  correspond to the oscillating components
with frequency around $2k_0(t)$ in real space, yielding an
interference pattern. The Gaussian blurring of the peaks is due to
the fact that the expanding wave functions do not have the same
amplitude at each point. The height of the side peaks is, as in
the case of the plane waves, one fourth the height of the central
peak, yielding in the large time limit an interference pattern
$\sim \cos^2k_0(t)x$.

\begin{figure}
\centering \subfigure{
\includegraphics[width=0.5\textwidth]{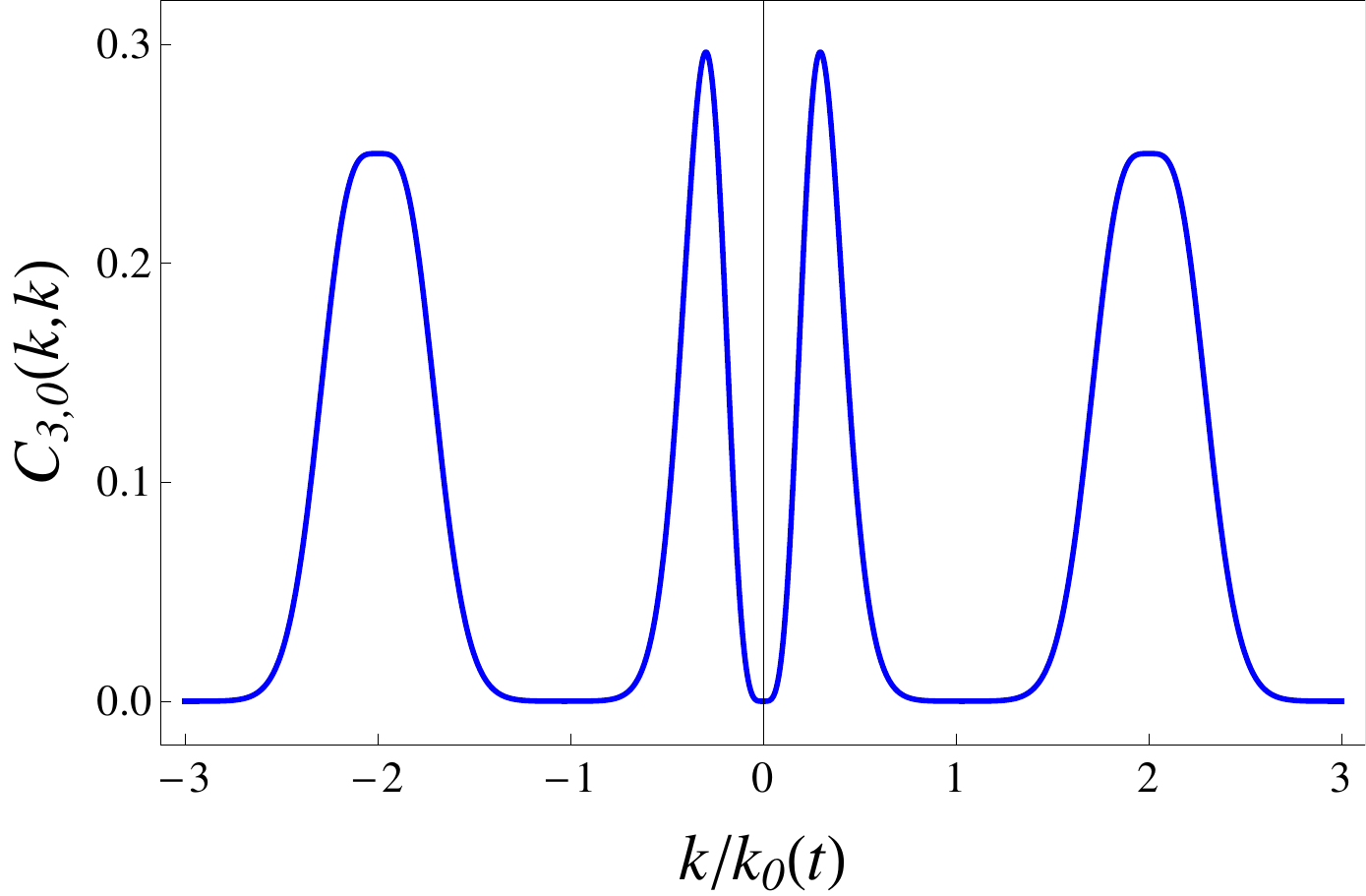}}
\subfigure{
\includegraphics[width=0.5\textwidth]{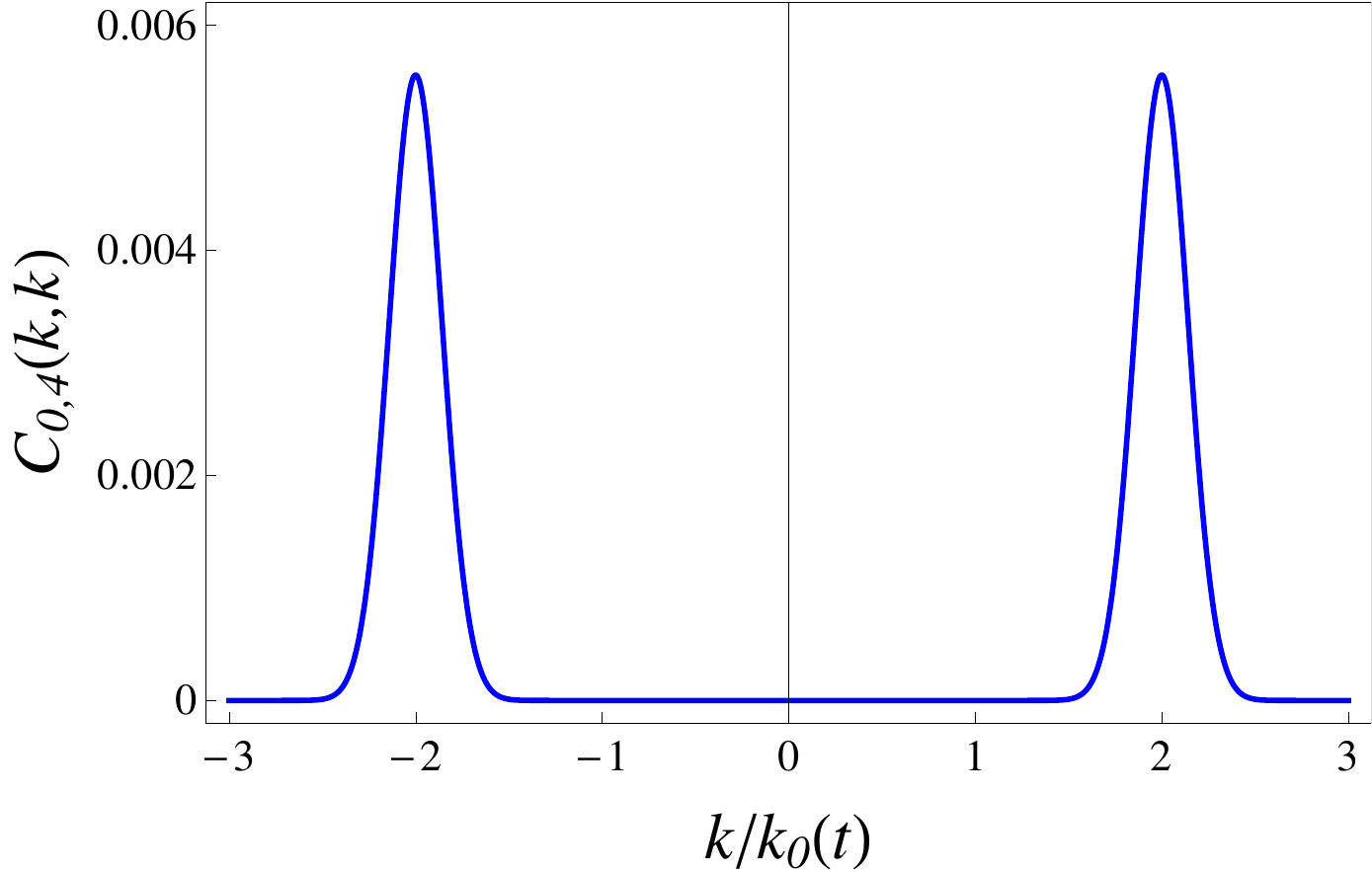}}
\caption{Dominant contributions to the variance of $R(k)$ in the
asymptotic regime, when $n$ is (a) much smaller, and (b) much
larger than $N^{3/4}$. Notice that the variances are in both cases
peaked around $k=\pm2k_0(t)$.}\label{gaussvar}
\end{figure}

Typicality will consist in proving that this outcome is valid for
the vast majority of wave functions of the condensate. In other
words, in most experimental runs, each characterized by a given
wave function of the condensate, one will observe an interference
pattern with fringes $\sim \cos^2k_0(t)x$. Let us therefore check
that the fluctuations of $R(k)$ around its average (\ref{rho2kgauss})
are small even in the large $n$ regime. Observe that, for large times,
the overlap between the functions $\mathcal{F}[\psi_a^*\psi_b]$,
$\mathcal{F}[\psi_b^*\psi_a]$, and $\tilde{\rho}_{a,b}$ is very
small, being $\text{O}(e^{-\alpha^2})$. Thus, the contributions
arising from the products $\mathcal{F}[\psi_a^*\psi_b](k)
\mathcal{F}[\psi_b^*\psi_a](k)$ and
$\mathcal{F}[\psi_{a,b}^*\psi_{b,a}](k) \tilde{\rho}_{a,b}(\pm k)$
can be consistently neglected, being as small as the overlap
between the two expanding Gaussian wave packets. This yields
\begin{equation}
(\delta R)^2 (k,k';t) \simeq \frac{n^3}{180} \left( e^{-
\frac{t^2}{2} [k + 2 k_0(t)]^2} + e^{- \frac{t^2}{2} [k - 2
k_0(t)]^2} \right) \left( e^{- \frac{t^2}{2} [k' + 2 k_0(t)]^2} +
e^{- \frac{t^2}{2} [k' - 2 k_0(t)]^2} \right) + \text{O}(N^2).
\end{equation}
In the light of the previous observation, we can compute the
$F_{i,j}$ functions in (\ref{rho4k}) and finally obtain the
statistical average of quantum covariance
\begin{equation}
\overline{(\Delta R)^2} (k,k';t) \simeq C_{3,0}(k,k') N^3 +
C_{1,2}(k,k') N n^2 + C_{0,4}(k,k') n^4 + C_{0,3}(k,k') n^3 +
\text{O}(N^2),
\end{equation}
with
\begin{eqnarray}
C_{3,0}(k,k') & = & 8 e^{- \frac{t^2}{2} (k^2 + {k'}^2)} \sinh^2
\!\left( \frac{t^2 k k'}{4} \right) \nonumber \\ &&{} - \frac{1}{4}
\left( e^{- \frac{t^2}{2} [k + 2 k_0(t)]^2} + e^{- \frac{t^2}{2}
[k - 2 k_0(t)]^2} \right) \left( e^{- \frac{t^2}{2} [k' + 2
k_0(t)]^2} + e^{- \frac{t^2}{2} [k' - 2 k_0(t)]^2} \right)
\nonumber \\ & &{}+ 2 e^{-\frac{t^2}{2} k^2} \left[
e^{-\frac{t^2}{2} [k' + 2 k_0(t)]^2} \sinh^2 \!\left( \frac{t^2 k
[k' + 2k_0(t)]}{4} \right) + e^{-\frac{t^2}{2} [k' - 2 k_0(t)]^2}
\sinh^2 \!\left( \frac{t^2 k [k' - 2k_0(t)]}{4} \right) \right] \nonumber \\
& &{}+ 2 e^{-\frac{t^2}{2} {k'}^2} \left[ e^{-\frac{t^2}{2} [k + 2
k_0(t)]^2} \sinh^2 \!\left( \frac{t^2 k' [k + 2k_0(t)]}{4} \right)
+ e^{-\frac{t^2}{2} [k - 2 k_0(t)]^2} \sinh^2 \!\left( \frac{t^2
k' [k - 2k_0(t)]}{4} \right) \right] \nonumber \\ &&{} +
\frac{1}{2}e^{-\frac{t^2}{2} (k - k')^2} \left(  e^{-\frac{t^2}{2}
[k + 2k_0(t)] [k' + 2k_0(t)]} + e^{-\frac{t^2}{2} [k - 2k_0(t)][k'
- 2k_0(t)]} \right) \nonumber \\ & &{}+
\frac{1}{2}e^{-\frac{t^2}{2} (k + k')^2} \left( e^{\frac{t^2}{2}
[k - 2k_0(t)] [k' + 2k_0(t)]} + e^{\frac{t^2}{2} [k + 2k_0(t)] [k'
- 2k_0(t)]} \right), \\ C_{1,2}(k,k') & = & \frac{1}{3} \left[ 8
e^{- \frac{t^2}{2} (k^2 + {k'}^2)} \sinh^2\! \left( \frac{t^2 k
k'}{4} \right) - C_{3,0}(k,k') \right], \\ C_{0,4}(k,k') & = &
\frac{1}{180} \left( e^{- \frac{t^2}{2} [k + 2 k_0(t)]^2} + e^{-
\frac{t^2}{2} [k - 2 k_0(t)]^2} \right) \left( e^{- \frac{t^2}{2}
[k' + 2 k_0(t)]^2} + e^{- \frac{t^2}{2} [k' - 2 k_0(t)]^2} \right)
= - C_{0,3}(k,k').
\end{eqnarray}
Also in this case,
fluctuations are small when $n=\text{o}(N)$.
This proves our thesis: interference is
typical and occurs for the overwhelming majority of wave functions
of the condensate.
One can again
distinguish two regimes, with threshold at $n=\text{O}(N^{3/4})$.
Figure \ref{gaussvar} displays the dominant contributions
$C_{3,0}$ and $C_{0,4}$ in the two regimes.

\section{Conclusions}
\label{sec-concl}

The search for a quantum mechanical explanation of statistical
mechanics dates back to the founding fathers of quantum
theory~\cite{Schroedinger,vonNeumann}. In the last few years this
subject has been revived and several successful new results have
been obtained. For a review, see \cite{gogolin}. In Refs.\
\cite{Lloyd,Tasaki,Gemmer,Popescu1,Popescu2,Goldstein,Sugita,Reimann,ref:Rigol-Nature,Cho,SugiuraShimizu} a
justification for the applicability of the canonical ensemble was
given that does not rely on subjective additional randomness added
by hand, or on ensemble averages. The statistical behavior is
shown to be a direct consequence of a genuine quantum approach. In
particular, in Refs.\ \cite{Lloyd,Popescu1,Popescu2} it was shown
that, under general assumptions on the Hamiltonian, typically
random pure states of large quantum systems  locally look  as the
microcanonical state. These works are based on typicality
arguments and on the mathematical phenomenon of measure
concentration \cite{Ledoux}. Typicality has also been shown to be
important in many emerging phenomena in physics and other
sciences. An interesting application is e.g.\  on the structure of
entanglement and of local entropies in large quantum
systems~\cite{Winter,mixmatrix,matrix
reloaded,polarized}.

In this article we have shown that typicality arguments also apply
to one of the most basic quantum phenomena: interference. Two
parts of a Bose-Einstein condensate, that are first separated and
then let overlap, will (almost) always interfere, no matter what
their wave function looks like, even if the splitting process
 divides the condensate in two unbalanced parts. No decoherence mechanism has been invoked.
Different
regimes have been identified, as a function of the scaling of $n$
with $N$. The threshold is at  number fluctuations as large as $n\sim N^\alpha$ with $\alpha=3/4$, which for $N\simeq 5000$ atoms yields $n\simeq 600$ atoms. Interference remains observable even for such an uneven splitting process.

One of the objectives of our future research will be to analyze
some recent experiments performed in Vienna
\cite{Schmied,Schmied2,Schmied3,Schmied4,Schmied5,Paraoanu} that bring to
light other interesting aspects of the interference of BECs, and
in particular the onset to decoherence, viewed as the
randomization of the relative phase as a function of the time
elapsed after the splitting process.

\appendix

\section{}

In this Appendix we give closed and general forms for the
$F_{i,j}$ functions appearing in (\ref{deltarho}). Let us
first introduce some shorthand for the combinations of Fourier
transforms that frequently appear in $F_{i,j}$:
\begin{eqnarray}
I_{ab}(\bm{k}) & = & \tilde{\rho}_a (\bm{k}) + \tilde{\rho}_b
(\bm{k}), \\ F_{ab}(\bm{k}_1,\bm{k}_2) & = &
\mathcal{F}[\psi_a^*\psi_b](\bm{k}_1)
\mathcal{F}[\psi_b^*\psi_a](\bm{k}_2) +
\mathcal{F}[\psi_b^*\psi_a](\bm{k}_1)
\mathcal{F}[\psi_a^*\psi_b](\bm{k}_2), \\
G_{ab}(\bm{k}_1,\bm{k}_2) & = & F_{ab}(\bm{k}_1,\bm{k}_2-\bm{k}_1)
+ F_{ab}(-\bm{k}_1,\bm{k}_2+\bm{k}_1),
\\ S_{ab}(\bm{k}_1,\bm{k}_2) & = & \tilde{\rho}_a (\bm{k}_1)
\tilde{\rho}_a (\bm{k}_2) +
\tilde{\rho}_b (\bm{k}_1) \tilde{\rho}_b (\bm{k}_2), \\
T_{ab}(\bm{k}_1,\bm{k}_2) & = & \tilde{\rho}_a (\bm{k}_1)
\tilde{\rho}_b (\bm{k}_2) + \tilde{\rho}_b (\bm{k}_1)
\tilde{\rho}_a (\bm{k}_2), \\ R_{ab}(\bm{k}) & = &
T_{ab}(\bm{k},-\bm{k}) + F_{ab}(\bm{k},-\bm{k}), \\
U_{ab}(\bm{k}_1,\bm{k}_2) & = & \tilde{\rho}_a(\bm{k}_1)
\tilde{\rho}_a(\bm{k}_2) \tilde{\rho}_b(\bm{k}_1+\bm{k}_2) +
\tilde{\rho}_b(\bm{k}_1) \tilde{\rho}_b(\bm{k}_2)
\tilde{\rho}_a(\bm{k}_1+\bm{k}_2), \\ V_{ab}(\bm{k}_1,\bm{k}_2) &
= & \tilde{\rho}_a(\bm{k}_1) \tilde{\rho}_a(\bm{k}_2)
\tilde{\rho}_a(\bm{k}_1+\bm{k}_2) + \tilde{\rho}_b(\bm{k}_1)
\tilde{\rho}_b(\bm{k}_2) \tilde{\rho}_b(\bm{k}_1+\bm{k}_2).
\end{eqnarray}
The functions which determine the fluctuations around the average
value of $|\tilde{\rho}(\bm{k})|^2$ read
\begin{eqnarray}
F_{4,0}(\bm{k},\bm{k}') & = & |\tilde{\rho}_a(\bm{k})
\tilde{\rho}_a(\bm{k}')|^2 + |\tilde{\rho}_b(\bm{k})
\tilde{\rho}_b(\bm{k}')|^2, \\ F_{3,1}(\bm{k},\bm{k}') & = &
R_{ab}(\bm{k}) S_{ab}(\bm{k}',-\bm{k}') + S_{ab}(\bm{k},-\bm{k})
R_{ab}(\bm{k}') + S_{ab}(\bm{k},\bm{k}') F_{ab}(-\bm{k},-\bm{k}')
\nonumber \\ & + & S_{ab}(-\bm{k},\bm{k}') F_{ab}(\bm{k},-\bm{k}')
+ S_{ab}(\bm{k},-\bm{k}') F_{ab}(-\bm{k},\bm{k}') +
S_{ab}(-\bm{k},-\bm{k}') F_{ab}(\bm{k},\bm{k}'), \\
F_{2,2}(\bm{k},\bm{k}') & = & |\tilde{\rho}_a(\bm{k})
\tilde{\rho}_b(\bm{k}')|^2 + |\tilde{\rho}_b(\bm{k})
\tilde{\rho}_a(\bm{k}')|^2 + R_{ab}(\bm{k}) R_{ab}(\bm{k}') +
T_{ab}(\bm{k},\bm{k}') F_{ab}(-\bm{k},-\bm{k}') +
T_{ab}(\bm{k},-\bm{k}') F_{ab}(-\bm{k},\bm{k}') \nonumber \\ & + &
T_{ab}(-\bm{k},\bm{k}') F_{ab}(\bm{k},-\bm{k}') +
T_{ab}(-\bm{k},-\bm{k}') F_{ab}(\bm{k},\bm{k}') +
\mathcal{F}[\psi_a^*\psi_b](\bm{k})
\mathcal{F}[\psi_a^*\psi_b](-\bm{k})
\mathcal{F}[\psi_b^*\psi_a](\bm{k}')
\mathcal{F}[\psi_b^*\psi_a](-\bm{k}') \nonumber \\ & + &
\mathcal{F}[\psi_b^*\psi_a](\bm{k})
\mathcal{F}[\psi_b^*\psi_a](-\bm{k})
\mathcal{F}[\psi_a^*\psi_b](\bm{k}')
\mathcal{F}[\psi_a^*\psi_b](-\bm{k}'), \\ F_{3,0}(\bm{k},\bm{k}')
& = & V_{ab}(\bm{k},\bm{k}') + V_{ab}(\bm{k},-\bm{k}') +
V_{ab}(-\bm{k},\bm{k}') + V_{ab}(-\bm{k},-\bm{k}'), \\
F_{2,1}(\bm{k},\bm{k}') & = & I_{ab}(\bm{k})
G_{ab}(\bm{k}',-\bm{k}) + I_{ab}(-\bm{k}) G_{ab}(\bm{k}',\bm{k}) +
I_{ab}(\bm{k}') G_{ab}(\bm{k},-\bm{k}') + I_{ab}(-\bm{k}')
G_{ab}(\bm{k},\bm{k}') \nonumber \\ & + & I_{ab}(\bm{k}+\bm{k}')
R_{ab}(-\bm{k},-\bm{k}') + I_{ab}(\bm{k}-\bm{k}')
R_{ab}(\bm{k},-\bm{k}') + I_{ab}(\bm{k}'-\bm{k})
R_{ab}(-\bm{k},\bm{k}') + I_{ab}(-\bm{k}-\bm{k}')
R_{ab}(\bm{k},\bm{k}') \nonumber \\ & + & U_{ab}(\bm{k},\bm{k}') +
U_{ab}(\bm{k},-\bm{k}') + U_{ab}(-\bm{k},\bm{k}') +
U_{ab}(-\bm{k},-\bm{k}').
\end{eqnarray}
Notice that the $F_{i,j}$ functions with $i+j=4$ come from the
eight-order term in the field operators appearing in
(\ref{rho4k}), while the ones with $i+j=3$ come from the
sixth-order terms under the integral in the same expression.

\end{document}